\documentclass[aps,prb,titlepage,superscriptaddress,twocolumn,showpacs,floatfix]{revtex4}

\usepackage{graphicx}
\usepackage{amsmath}
\usepackage[T1]{fontenc}
\usepackage[latin1]{inputenc}
\usepackage{times}
\usepackage[normalem]{ulem}
\usepackage{setspace}
  \makeatletter
  \def\@dotsep{4.5}
  \makeatother
\newlength{\myVSpace}
\newcommand{\ket}[1]{\left| #1 \right\rangle}

\newcommand{\be}{\begin{equation}}
\newcommand{\ee}{\end{equation}}
\newcommand{\ba}{\begin{eqnarray}}
\newcommand{\ea}{\end{eqnarray}}
\begin{document}

\title{Suppression of exciton dephasing in quantum dots through ultrafast multipulse control}
\author{Thomas E. Hodgson}
\affiliation{Department of Physics, University of York, Heslington, York,
YO10 5DD, United Kingdom}
\author{Lorenza Viola}
\affiliation{\mbox{Department of Physics and Astronomy, 6127 Wilder Laboratory, Dartmouth College, Hanover, New Hampshire 03755, USA}}
\author{Irene D'Amico}
\affiliation{Department of Physics, University of York, Heslington, York,
YO10 5DD, United Kingdom}

\pacs{03.67.Lx,73.21.La,81.07.Ta}
\begin{abstract}

We investigate the usefulness and viability of the scheme developed by Viola and Lloyd [Phys. Rev. A {\bf 58}, 2733 (1998)] to control dephasing in the context of exciton-based quantum computation with self-assembled quantum dots.  We demonstrate that optical coherence of a confined exciton qubit exposed to phonon-induced dephasing can be substantially enhanced through the application of a simple periodic sequence of control pulses. The shape of the quantum dot has a significant effect on the dephasing properties. Remarkably, we find that quantum dots with parameters optimized for implementing quantum computation are among the most susceptible to dephasing, yet periodic decoupling is most efficient for exactly that type of dot. We also show that the presence of an electric field, which is a necessary ingredient for many exciton-based quantum computing schemes, may further increase the control efficiency. Our results suggest that dynamical decoupling may be a method of choice for robust storage of exciton qubits during idle stages of quantum algorithms.

\end{abstract}

\date{\today}
\maketitle

\section{Introduction}

The quest for practical implementations of quantum information processing (QIP) \cite{neilsenchang} has resulted in a wide variety of proposed quantum computing architectures, based on virtually all kinds of different physical systems \cite{qipexp}.  Thanks to continuous advances in nanoscale design and manufacturing of low-dimensional structures, solid-state platforms promise, in particular, a high level of scalability and large-scale integration. Among the various proposals which exist to date, schemes which rely on {\em charge} degrees of freedom of carriers ({\em excitons}) confined in a self-assembled semiconductor quantum dot (QD) \cite{PhysToday} are especially appealing in view of current ultrafast spectroscopic capabilities, which make it possible to access sub-picosecond gating times by means of suitable all-optical control techniques \cite{chapter,carlo}. 

In practice, such advantages are hindered by the rate at which quantum information stored in exciton qubits and/or exciton ancillary states is irreversibly lost due to dephasing -- as characterized by a typical time scale $T_2$. For QDs to be employed in quantum information devices, decoherence processes, which involve energy exchange with the host crystal are highly suppressed due to the large splitting between quantized energy levels. Thus, the dominant contribution to $T_2$ arises from energy-conserving transitions which do not change the occupations of the logical states, but result in {\em pure dephasing} and unrecoverable loss of phase information. While typical $T_1$ times i.e. finite qubit lifetime via exciton recombination, for single QDs are in the nanosecond range, it 
has been shown theoretically \cite{1stkuhn,kuhnpulses,kuhnirene} and experimentally \cite{kuhnexp} that substantial dephasing can occur in much shorter time scales (one or two picoseconds at temperatures of a few K), mainly due to coupling with acoustic phonons.  

Borrowing inspiration from coherent averaging techniques in high-resolution nuclear magnetic resonance spectroscopy \cite{NMR}, dynamical decoupling (DD) methods have recently emerged in QIP as a versatile strategy for counteracting decoherence from non-Markovian environments, in a way which is substantially less resource-demanding than traditional quantum error-correcting codes. In its simplest form, DD is implemented by subjecting the system to periodic sequences of fast and strong (so-called {\em bang-bang}) control operations drawn from a basic repertoire, so that interactions whose correlation times are long compared to the control time scale are effectively averaged out to finite accuracy \cite{lorenza,PRLdd1}.  In particular, a number of analytical and numerical studies have established the potential for DD schemes based on high-level concatenated \cite{kaveh} or randomized \cite{lea,kern} design to extend the coherence time of spin degrees of freedom in gate-defined semiconductor QDs by (at least) two orders of magnitude \cite{souza,witzel,wen,sham}. Remarkably, simple DD protocols consisting of a single or multiple spin-echoes have been experimentally demonstrated in (gate-defined) double QD devices \cite{petta} and rare-earth solid-state centers \cite{sellars}, respectively. In this paper we will consider {\em self-assembled} semiconductor QDs.

The usefulness of a suitable burst of ultrashort pulses as a tool to reduce the initial optical polarization dephasing in this type of QD has been demonstrated by Axt {\em et al}~\cite{kuhncontrol}.  In this work, we investigate the effectiveness of multipulse DD toward both mitigating the initial coherence decay and ensuring enhanced coherence preservation over a desired storage time scale, with emphasis on QIP applications.  Our analysis has two main implications: first, we show that in spite of being most prone to dephasing, QDs which are optimized to support quantum computation are also most efficiently stabilized by DD.  Furthermore, thanks to the fact that the exciton spectral density is down-shifted in frequency by the application of a static electric field, DD performance are further improved for a biased QD as required for most computational applications\cite{chapter,irene}. 

The content of the paper is organized as follows.  The next section provides an in-depth analysis of the free dephasing dynamics of an exciton qubit in the absence of applied control. After laying out the relevant QD model in Sec. IIA, special emphasis is devoted to isolate and analyze the contributions of the piezoelectric and deformation couplings to the exciton spectral density (Sec. IIB) for various dot shapes, as well as to quantify the influence of the spectral density details on the dephasing behavior.  In Sec. III, we summarize our results on the controlled dephasing dynamics, and characterize the dependence of DD performance upon different system and control parameters -- including dot shape, temperature, separation between pulses, and application of an external bias field.  In particular, DD performance for robust exciton storage in QIP applications are highlighted 
in Sec. IIIC, IIID, respectively.  Concluding remarks follow in Sec. IV.

\section{Dephasing dynamics of confined excitons}

\subsection{QD model Hamiltonian}

We consider a GaAs/AlAs QD with either $0$ or $1$ ground state exciton\cite{kuhnirene}, which corresponds to the qubit logical states $\ket{0}$ or $\ket{1}$ respectively. The Hamiltonian of this two-level excitonic system interacting with the phonon modes of the lattice is given by
\begin{equation}
H=E_{exc}c^{\dagger}c+\hbar \hspace*{-0.6mm}\sum_{j,{\bf k}} \omega_j( {\bf k} )b^{\dagger}_{j,{\bf k}}b_{j,\bf{k}}+\hbar \,c^{\dagger}c
\hspace*{-0.4mm} \sum_{j,\bf{k}} (g^*_{j,\bf{k}}b^{\dagger}_{j,\bf{k}}
+ H.c.),
\label{hamiltonian}
\end{equation}
where $E_{exc}$ is the energy of the ground state exciton relative to the crystal ground state, $c^{\dagger} (c)$ are creation (annihilation) operators for an exciton, $b^{\dagger}_{j,{\bf k}} (b_{j,\bf{k}})$ are bosonic creation (annihilation) operators for a phonon of mode $j$, wave vector ${\bf k}$, and angular frequency $\omega_j ({\bf k})$, and $g_{j,\bf{k}}$ is the coupling between the exciton and a phonon of mode $j,{\bf k}$, respectively.

We assume that the exciton and the phonon bath are initially uncorrelated, with the phonon bosonic reservoir being in thermal equilibrium at temperature $T$. As time evolves, the qubit becomes entangled with the environment and the off-diagonal elements of the exciton density matrix evaluated at time $t$ in the interaction picture with respect to the free system and bath Hamiltonians are given by~\cite{lorenza,1stkuhn}
\begin{equation}
\rho_{01}(t)=\rho_{10}^*(t)=\rho_{01}(0)e^{-\Gamma (t)}, \label{dens}
\end{equation}
with
\begin{equation}
\Gamma(t)=\int_{0}^{\infty} {d\omega}\, \frac{I(\omega)}{\omega^2}
\coth\Big(\frac{\hbar\omega}{k_B T}\Big)(1-\cos(\omega t)).
\label{Gam1}
\end{equation}
Here, $k_B$ is the Boltzmann constant, and the spectral function
\begin{equation}
I(\omega)= \sum_{j,{\bf k}}\delta(\omega-\omega_j({\bf k}))|g_{j,\bf{k}}|^2 \label{Iom}
\end{equation}
describes the interaction of the qubit with a phonon of frequency $\omega_j(\bf{k})$. For an exciton qubit, $\Gamma(t)$ is formally a factor of 4 smaller than for a spin coupled to a bath of harmonic oscillators as described in Ref. \onlinecite{lorenza}. This is due to the different Hamiltonians. For the exciton qubit the coupling to the environment only occurs when the qubit is in the logical state $\ket{1}$ (exciton present), while a spin qubit couples to the environment in both logical states. 

In order to maximize coherence, we consider sufficiently low temperatures, so that the coupling to optical phonons may be safely neglected. For acoustic phonons, we assume the dispersion relation
$\omega_j({\bf k})=v_j|{\bf k}|$.
In addition, we assume a small lattice mismatch at the boundary between the QD and the barrier, and therefore approximate the coupling to the phonon bath as the bulk coupling modulated by the appropriate form factor~\cite{1stkuhn}:
\begin{equation}
g_{j,\bf{k}}=\int d^3{\bf r}_e d^3{\bf r}_h|\Psi({\bf r}_e, {\bf r}_h)|^2(G^e_{j,\bf{k}}e^{i{\bf k. r}_e}- G^h_{j,\bf{k}}e^{i{\bf k. r}_h}).
\end{equation}
Here, $\Psi({\bf r}_e, {\bf r}_h)$ is the exciton wave function and $G^{e/h}_{j,{\bf k}}$ is the bulk coupling of the single particle to the phonon bath,
\begin{equation}
G^{e/h}_{j,\bf{k}}=\frac{1}{\sqrt{2\varrho\hbar\omega_{j}({\bf k})V}}[kD^{e/h}_j+iM_j({\hat{\bf k}})].
\end{equation}
The index $j$ runs over the two transverse and the longitudinal modes, 
$\varrho$ is the density of the QD, and $V$ is a normalization volume. Excitons in bulk semiconductor couple to longitudinal acoustic phonons through a deformation potential coupling $D^{e/h}_j$, and to all the modes considered through a piezoelectric potential coupling $M_j$. 
For zincblende crystals such as GaAs/AlAs,
\begin{equation}
M_j({{\bf \hat{k}})=\frac{2e_{14}e}{\epsilon_s\epsilon_0}}
\Big(\hat{k}_x\hat{k}_y\hat{\xi}_z^{(j)}+\hat{k}_y\hat{k}_z\hat{\xi}_x^{(j)}+\hat{k}_z\hat{k}_x\hat{\xi}_y^{(j)}\Big).
\end{equation}
In what follows, we shall approximate $|M_j|^2$ by its angular average~\cite{1stkuhn},
\begin{equation}
\frac{1}{4\pi}\int^{2\pi}_{0}d\phi\int^{\pi}_{0}d\theta \sin(\theta)M^2_j({\bf \hat{k}})=A_j\left(\frac{2ee_{14}}{\epsilon_s\epsilon_0}\right)^2,
\end{equation}
where $e$ is the electron charge, $e_{14}$ is the piezoelectric coefficient, and $\epsilon_s$, $\epsilon_0$ denote the relative permittivity of the material and of free space, respectively. $\hat{\xi}_{x/y/z}$ are unit vectors describing the polarization of mode ${\bf k}$, and $A_j$ are mode-dependent geometrical factors.

We will consider nanostructures in the {\em strong confinement} regime, so that Coulomb coupling may be neglected and excitons may be accurately modeled as products of Gaussian wave-packets within the single particle approximation. We assume that the QD is isotropic in the plane perpendicular to the growth direction. The excitonic wave-function is then
\begin{equation}
\Psi({\bf r}_e, {\bf r}_h)=\psi({\bf r}_{e})\psi({\bf r}_{h}),
\end{equation}
where
\begin{equation}
\psi({\bf r}_{e/h})=\frac{1}{(\lambda_{r_{e/h}}^2\lambda_{z_{e/h}}\pi^{\frac{3}{2}})^\frac{1}{2}}\exp\bigg(-\frac{r_{e/h}^2}{2\lambda_{r_{e/h}}^2}-\frac{z_{e/h}^2}{2\lambda_{z_{e/h}}^2}\bigg),
\end{equation}
and $2\lambda_{z_{e/h}}$ and $2\lambda_{r_{e/h}}$ are the characteristic widths of the electron/hole wavefunction in the $z$ (growth) direction, and in the plane $r=\sqrt{x^2+y^2}$ respectively, and are related in the usual way to the characteristic frequencies of the parabolic confining potential in the appropriate direction. For later reference, we define a characteristic volume of the QD,
\begin{equation}
 V_{\psi}\equiv8\lambda_{r_e}^2\lambda_{z_e}.
\end{equation}

For the exciton-phonon system, Eq.~(\ref{Iom}) can be written as
\begin{equation}
I_{exc}(\omega)=I_e(\omega)+I_h(\omega)-2I_{eh}(\omega),
\end{equation}
where $I_{e/h}(\omega)$ is the spectral density of the electron/hole, and $I_{eh}(\omega)$ modifies the spectral density due to the electron hole interference. Each term has the following form:
\begin{equation}
I_{s}(\omega)\sim\sum_j\omega|G_{j,s}'(\omega)|^2 f_{s}\Big(\frac{\omega \xi_{s}}{v_j}\Big)\exp\Big(-\frac{\lambda_{r_{s}}^2 \omega^2}{2v_j^2}\Big),
\label{Ianalytical}
\end{equation}
where the index $s$ runs over $e$, $h$, and $eh$, 
\begin{equation}
\xi_{s}^2=\frac{|\lambda_{z_{s}}^2-\lambda_{r_{s}}^2|}{2},
\end{equation}
and
\begin{equation}
\lambda_{(r/z)_{eh}}^2=\frac{\lambda_{(r/z)_e}^2+\lambda_{(r/z)_h}^2}{2}.
\end{equation}
The form of $f_s(x)$ depends on the dot shape. For $\lambda_{r_s}>\lambda_{z_s}$ ({\em oblate}),
\begin{equation}
f_{s}^{obl}(x)=\frac{i \operatorname{erf}(ix)}{x},
\end{equation}
where $\operatorname{erf}(x)$ is the error function; for $\lambda_{r_s}<\lambda_{z_s}$ ({\em prolate}),
\begin{equation}
f_{s}^{pro}(x)=\frac{\operatorname{erf}(x)}{x},
\end{equation}
and for $\lambda_{r_s}=\lambda_{z_s}$ ({\em spherical})
\begin{equation}
f_{s}^{sph}(x)=\frac{2}{\sqrt{\pi}}.
\end{equation}
In Eq.~(\ref{Ianalytical}), $|G_{j,s}'(\omega)|^2$ includes terms due to both the deformation potential and the piezoelectric coupling, that is, 
\begin{equation}
|G_{j,s}'(\omega)|^2=\frac{\omega^2}{v_j^2}D_j^{s^2}+|M_j|^2, 
\label{Gprime}
\end{equation}
where
$(D^{eh})^2=D^eD^h$.
This allows the spectral density to be separated into the sum of the spectral densities related to each of the two coupling mechanisms,
\begin{equation}
I(\omega) =I^{def}(\omega)+I^{piez}(\omega),
\end{equation}
and the total evolution in Eq.~(\ref{dens}) to be expressed as a product of contributions from the deformation potential and piezoelectric interactions,
\begin{equation}
e^{-\Gamma(t)}=e^{-\Gamma^{def}(t)}e^{-\Gamma^{piez}(t)}.
\end{equation}

\subsection{Single particle and excitonic spectral density}

The dephasing dynamics of the exciton qubit is determined by the exciton spectral density. In general, $I(\omega)$ is positive, goes to zero as $\omega\rightarrow 0$, and becomes negligible for large $\omega$. Aside the material parameters, the main factor influencing the spectral density is the {\em shape} of the QD~\cite{1stkuhn}, which enters our calculations through the values of $\lambda_{(r/z)_e}$ and $\lambda_{(r/z)_h}$. 

\begin{figure}[t]
\includegraphics*[width=\linewidth]{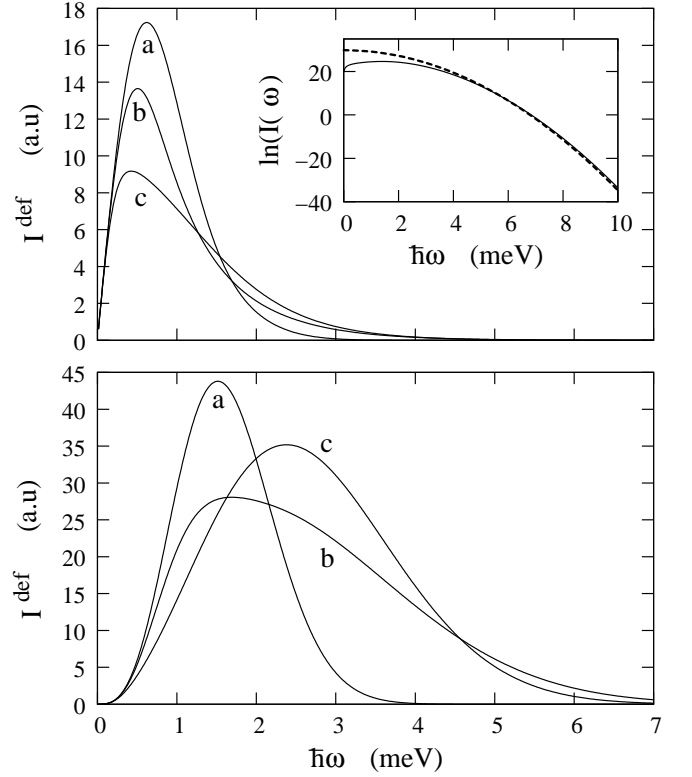}
\caption{Piezoelectric $I_e^{piez}(\omega)$ (top) and deformation $I_e^{def}(\omega)$ (bottom) contributions to the electron spectral density for the cases of: a) spherical QD ($\lambda_{r_e}=3.84$~nm), b)  oblate QD 
($\lambda_{r_e}=6$~nm), and c) prolate QD ($\lambda_{r_e}=2$~nm). Inset: 
$\ln I(\omega)$ for a single electron with $\lambda_{r_e}=3.84$~nm (solid line). The dashed line is the parabolic fit with the curve $(29.7-2.79$x
$10^{-25}\omega^2)$, see Eq.~(\ref{cutofflog}).}
\label{electronpiezspect}
\end{figure}

To understand the system at hand, we will first consider the behavior of the spectral density for a single trapped particle. Throughout this paper we will consider QDs with $V_{\psi_e}=455.28$~nm$^3$. For a single particle, for instance an electron, the maximum of the spectral density is greatest for a spherical QD. For oblate or prolate QDs with the same characteristic volume $V_{\psi}$, the spectral density has a lower maximum, but extends to higher frequencies. The top panel of Fig. \ref{electronpiezspect} shows the piezoelectric spectral densities for a single electron, $I^{piez}_e$, for different dot shapes.
As the dot becomes more asymmetric, the interplay between $f_e(\omega\xi_e/v_j)$ and the Gaussian term in Eq. (\ref{Ianalytical}) reduces the maximum of $I^{piez}_e(\omega)$, whilst increasing $I^{piez}_e(\omega)$ at high frequencies. 

For the deformation potential coupling, the single particle spectral densities follow a similar trend, however the extra $\omega^2$ term in the expression for $I_e^{def}(\omega)$ (see Eq. (\ref{Gprime})) results in larger $I_e^{def}(\omega)$ at high frequencies. The bottom panel of Fig. \ref{electronpiezspect} shows the deformation potential spectral densities for a single electron in QDs with different shapes.

We can define an upper cut-off frequency $\omega_c$, such that $I(\omega)$ is negligible for $\omega \gg \omega_c$. The QD spectral density is composed of a sum of terms, each containing a Gaussian function. The high frequency behaviour of the overall function is therefore very similar to a Gaussian, thus we may write
\begin{equation}
I(\omega)\approx F(\omega)e^{-\frac{\omega^2}{\omega_c^2}},  
\label{cutoff}
\end{equation}
and treat $\omega_c$ as the cut-off frequency. In order to obtain a value for $\omega_c$, we consider
\begin{equation}
\ln I(\omega)= \ln F(\omega)-\frac{\omega^2}{\omega_c^2},  
\label{cutofflog}
\end{equation}
and fit this form to our numerical results. The inset of Fig. \ref{electronpiezspect} shows $\ln I(\omega)$ for a single electron in a spherical QD ($\lambda_{r_e}=3.84$~nm), along with fitted curve $29.7-2.79$x$10^{-25}\omega^2$ which leads to $\omega_c=1.89$x$10^{12}$~rads$^{-1}$ ($\hbar\omega_c=1.25$~meV). 

For an exciton in a self-assembled QD, the electron and hole confinement in the $z$ (growth) direction is due to the band-offset between the barrier and dot layers. Because it is then mainly determined by the thickness of these layers, $\lambda_{z_e}\approx\lambda_{z_h}$. The radial in-plane confinement, however, may be different for electron and hole, and is accurately modeled by harmonic potentials.  This leads to
\begin{equation}
\lambda_{r_h}=\sqrt{\frac{m_e^*\omega_e}{m_h^*\omega_h}}\lambda_{r_e}, 
\label{m}
\end{equation}
where $\omega_{e/h}$ is the characteristic frequency of the confining potential for the electron/hole, and $m^*_{e/h}$ is the electron/hole effective mass. Throughout this paper, we shall discuss GaAs QDs~\cite{irene}, the material parameters of which are listed in Table \ref{parameters}. Because of Eq.~(\ref{m}), the electron and hole wave-functions are not simultaneously spherical for a given QD. There are three possible regimes for the exciton wave-function. For a heavily prolate QD, both the electron and hole wave-functions are prolate ({\em prolate-prolate regime}), similarly for heavily oblate QDs the wave-functions are both oblate ({\em oblate-oblate regime}). In between these two situations, the electron and hole will stretch in different directions ({\em oblate-prolate regime}). The most symmetric configuration in the oblate-prolate regime occurs when $\xi_e=\xi_h$, corresponding to the following radial confinement:
\begin{equation}
{\bar\lambda}_{r_e}=\left(\frac{V_{\psi}^e}{4\sqrt{2(1+\frac{m_e^*\omega_e}{m_h^*\omega_h}})}\right)^\frac{1}{3}.
\end{equation}
In the results plotted throughout the paper, we shall assume $\hbar\omega_e/\hbar\omega_h=1.25$, which corresponds to the parameters for the optimal dot for quantum computation discussed in Ref. \onlinecite{irene}.

\begin{table}[ht]
\begin{center} {\footnotesize
\begin{tabular}{lcc}
\hline 
 \hline
Static dielectric constant & $\epsilon_s$ & $12.53$\\
Longitudinal sound velocity $(m/s)$ &  $v_{L}$ & $5110$\\
Transverse sound velocity $(m/s)$ &  $v_{T}$ & $3340$\\
Density $(g/cm^3)$ & $\varrho$ & $5.37$\\
Electron deformation potential $(eV)$ & $D_e$ & $7.0$\\
Hole deformation potential $(eV)$ & $D_h$ & $-3.5$\\
Piezoelectric constant $(C/m^2)$ & ${e_{14}}$ & $0.16$\\
Effective electron mass $(m_0)$ & $m_e$ & $0.067$\\
Effective Hole mass $(m_0)$ & $m_h$ & $0.34$\\
Longitudinal geometrical factor & $A_L$ & $3/35$\\
Transverse geometrical factor & $A_T$ & $1/21$\\
\hline
\end{tabular} }
\end{center}
\caption{Material parameters for GaAs QDs, from Ref. \onlinecite{kuhnirene}. $m_0$ denotes the free electron mass.}
\label{parameters}
\end{table}

\begin{figure}[b]
\includegraphics*[width=\linewidth]{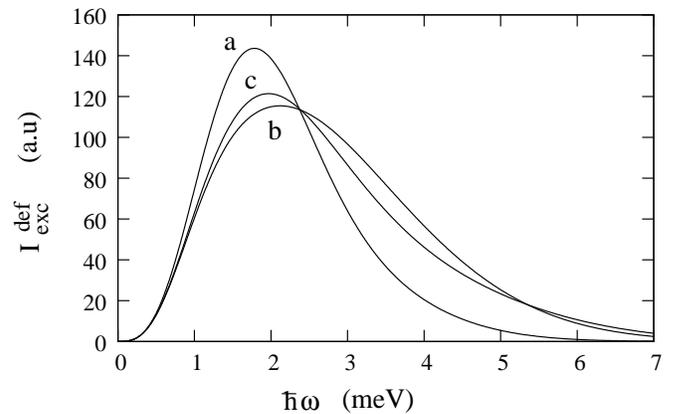} 
\caption{Excitonic deformation spectral density $I_{exc}^{def}(\omega)$ 
for the cases of: a) 
$\lambda_{r_e}=\bar\lambda_{r_e}=4.16$~nm, b) oblate-oblate regime 
($\lambda_{r_e}=6$~nm), and c) prolate-prolate regime ($\lambda_{r_e}=3$~nm).}
\label{deformexcitonI}
\end{figure}

For the excitonic system, there is an additional difference between the spectral density components due to piezoelectric and deformation potential couplings: for piezoelectric coupling, the interference term $I^{piez}_{eh}$ is positive, whereas $I^{def}_{eh}$ is negative. For deformation potential coupling, the interference term is of the same order of $I^{def}_e$ and $I^{def}_h$. As a consequence, $I_{exc}^{def}$ behaves in a similar way to the single particle spectral density in that it has a lower maximum in the prolate-prolate and oblate-oblate regimes compared to QDs with $\lambda_{r_e}\approx\bar\lambda_{r_e}$, however $I_{exc}^{def}$ will be greater for large 
$\omega$. 

Fig. \ref{deformexcitonI} shows the deformation potential spectral density $I_{exc}^{def}$ for different dot shapes.
Notice that at $\lambda_{r_e}=\bar\lambda_{r_e}$, the electron wave-function is as oblate as the hole wave-function is prolate; for deformation potential coupling, $I^{def}_{exc}$ behaves similarly for $\lambda_{r_e}>\bar\lambda_{r_e}$ as for $\lambda_{r_e}<\bar\lambda_{r_e}$. 

\begin{figure}[t]
\includegraphics*[width=\linewidth]{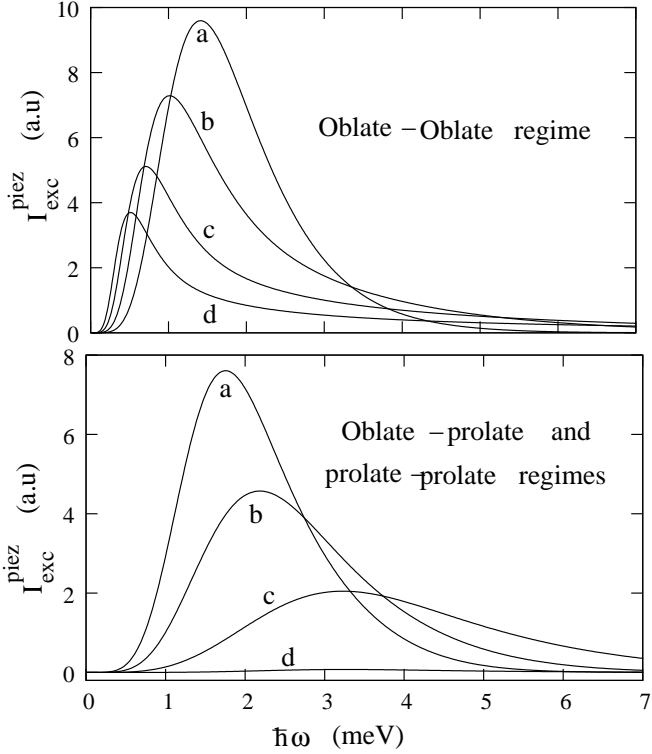} 
\caption{Top panel: Excitonic piezoelectric spectral density $I^{piez}_{exc}(\omega)$ in the oblate-oblate regime, for the cases of increasing $\lambda_{r_e}$. a) $\lambda_{r_e}=6$~nm, b)  $\lambda_{r_e}=9$~nm, c) $\lambda_{r_e}=13$~nm, and d) $\lambda_{r_e}=18$~nm. Bottom panel: $I^{piez}_{exc}(\omega)$ in the oblate-prolate and prolate-prolate regimes, respectively, for decreasing $\lambda_{r_e}$. 
a) $\lambda_{r_e}=4$~nm, b)  $\lambda_{r_e}=3$~nm, c) $\lambda_{r_e}=2$~nm, and d) $\lambda_{r_e}=1$~nm. }
\label{piezexcitonIlarger}
\end{figure}

For piezoelectric coupling, the different sign of the interference term 
$I^{piez}_{eh}$ leads to a very different behavior of the spectral density. First, if the electron and hole wave-functions were exactly the same, there would be no piezoelectric coupling, that is,  $I_{exc}^{piez}=0$. At low frequencies, $I^{piez}_e(\omega)\approx I^{piez}_h(\omega)$, and their sum cancels with the contribution from the interference term, leading to $I_{exc}^{piez}\sim 0$. The width of this depletion of the spectral density is greater for smaller $\lambda_{r_e}$. For dots in the oblate-oblate regime ($\lambda_{r_e}>4.85$~nm), as $\lambda_{r_e}$ is decreased the maximum of the piezoelectric exciton spectral density increases, however it also shifts towards higher frequency modes as shown in the top panel of Fig. \ref{piezexcitonIlarger}. As $\lambda_{r_e}$ decreases further, we first enter the oblate-prolate regime ($3.84$~nm$<\lambda_{r_e}<4.85$~nm) and then the prolate-prolate regime ($\lambda_{r_e}<3.84$~nm). As $\lambda_{r_e}$ reduces to even smaller values, the piezoelectric spectral density becomes small for all $\omega$. This is due to the type of confinement in the growth direction, which results in $\lambda_{z_e}=\lambda_{z_h}$. In this limit, $\xi_e\approx\xi_h$ and $I^{piez}_e+I^{piez}_h+I^{piez}_{eh}\approx0$ for all $\omega$. This behavior may be seen in the bottom panel of Fig. \ref{piezexcitonIlarger}, in which $I^{piez}_{exc}$ is plotted for values of decreasing $\lambda_{r_e}$. 
We shall denote the value of $\lambda_{r_e}$ which yields the largest maximum of $I^{piez}_{exc}$ as $\bar\lambda^{piez}_{r_e}$, with $\bar\lambda^{piez}_{r_e}\gtrsim \bar\lambda_{r_e}$. For $V_{\psi_e}=455.28$~nm$^3$, $\bar\lambda^{piez}_{r_e}=5.58$~nm. The discrepancy between $\bar\lambda_{r_e}$ and $\bar\lambda^{piez}_{r_e}$ is due to the constraint $\lambda_{z_e}=\lambda_{z_h}$.  

Again, we may follow the same method as in the single-electron case to define and estimate a spectral cut-off $\omega_c$. For an exciton with $\lambda_{r_e}=\bar\lambda_{r_e}=4.16$~nm, this yields $\omega_c=3.52$x$10^{12}$~rads$^{-1}$ ($\hbar\omega_c=2.32$~meV), i.e. $\omega_c$ is of the same order as the single particle case for a spherical QD.

\subsection{Effect of spectral density on exciton dephasing}

\begin{figure}[t]
\includegraphics*[width=\linewidth]{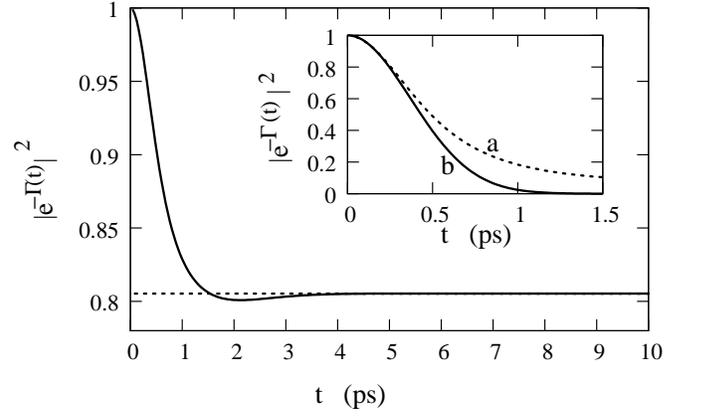}
\caption{Decoherence dynamics for $\lambda_{r_e}=6.16$~nm at 4K (solid line), and asymptotic decoherence value from Eq. (\ref{Gam3}) (dashed line). Inset: Decoherence dynamics for $\lambda_{r_e}=6.16$~nm at 77K, using: a) exact results, and b) the short-time approximation from Eq. 
(\ref{Gam4}). }
\label{examplefree}
\end{figure}

Insight into the dephasing properties may be obtained by examining the evolution of the off-diagonal density matrix elements with time, Eq.~(\ref{dens}). Physically, the coherence element $\rho_{01}(t)$ is proportional to the system optical polarization at time $t$, $ {\bf P}(t)$ (see e.g. Eq.~(9) in Ref.~\onlinecite{1stkuhn}). In order to emphasize this relation, we have chosen to monitor the quantity 
$|e^{-\Gamma (t)}|^2=|{\bf P}(t)|^2/|{\bf P}(t=0)|^2$. Fig.~\ref{examplefree} shows the decoherence factor $|\exp(-\Gamma(t))|^2$ for $\lambda_{r_e}=6.16$~nm at 4K. $|\exp(-\Gamma(t))|^2$ falls quickly from unity before saturating at a temperature- and shape- dependent value. For the examples in Fig. \ref{examplefree}, $|\exp(-\Gamma(\infty))|^2$ is non zero, but at higher temperatures and for differently shaped dots $\rho_{01}$ may fall to zero~\cite{1stkuhn}.

\begin{figure}[t]
\includegraphics*[width=\linewidth]{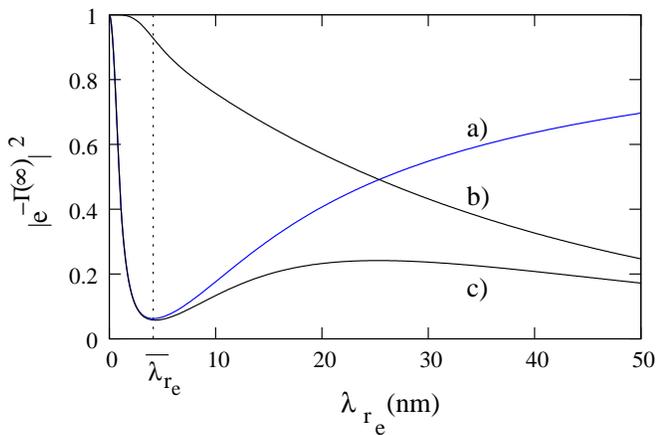}
\caption{Asymptotic decoherence factor $|\exp(-\Gamma(\infty))|^2$ versus 
$\lambda_{r_e}$ for: a) deformation potential coupling, b) piezoelectric coupling, and c) both effects included, for an exciton at $T=77$K.}\label{asymptot}
\end{figure}

More specifically, in the limit of $t\rightarrow\infty$, Eq. (\ref{Gam1}) becomes
\begin{equation}
\Gamma(\infty)=\int_{0}^{\infty}d\omega\frac{ I(\omega)}{\omega^2}
\coth\Big(\frac{\hbar\omega}{k_B T}\Big),
\label{Gam3}
\end{equation}
therefore the non-diagonal density matrix may saturate to a finite value if $I(\omega)\stackrel{\omega\to 0}{\sim} \omega^3$. This was discussed for specific spectral density functions in Refs.~\onlinecite{1stkuhn} and \onlinecite{palmer}. Saturation at a non-zero value does not occur for the spin-boson system with Ohmic spectral density analyzed in Ref. \onlinecite{lorenza}. Fig. \ref{asymptot} shows the asymptotic coherence value $|\exp(-\Gamma(\infty))|^2$ as a function of $\lambda_{r_e}$ for an exciton at 77 K in the presence of a) deformation potential coupling alone, b) piezoelectric coupling alone, and c) both effects included. For the piezoelectric interaction, $\Gamma^{piez}(\infty)\rightarrow 0$ as $\lambda_{r_e}\rightarrow 0$ due to the fact that $I_{exc}^{piez}\rightarrow0$ (Fig. \ref{piezexcitonIlarger}). As $\lambda_{r_e}$ increases, $\Gamma^{piez}(\infty)$ becomes large, reflecting the fact that the asymmetry between electron and hole wave-functions becomes relevant, therefore increasing the piezoelectric coupling.  For the deformation potential coupling, $\Gamma^{def}(\infty)$ is maximum very close to $\bar\lambda_{r_e}$.
It can be seen in Fig. \ref{asymptot} that in order to minimize the long-term decoherence, we should either consider very prolate QDs ($\lambda_{r_e} \ll \lambda_{z_e}$) -- for which piezoelectric coupling can be neglected -- or consider the local maximum of $|e^{-\Gamma(\infty)}|^2$ -- which occurs at $\Gamma^{piez}\approx\Gamma^{def}$, at $\lambda_{r_e}\approx25$~nm in the case of Fig. \ref{asymptot}. The results in Fig. \ref{asymptot} are calculated at a temperature of 77K. The magnitude of the dephasing can be reduced by considering lower temperatures.

Let us now consider the opposite time limit. The short-time decoherence behavior offers insight into how the spectral density affects the initial dephasing rate, that is, how fast $\exp(-\Gamma(t))$ falls from unity. For small $t$, Eq. (\ref{Gam1}) becomes
\begin{equation}
\Gamma(t)\approx\frac{t^2}{2}\int_{0}^{\infty}d\omega I(\omega)
\coth\Big(\frac{\hbar\omega}{k_B T}\Big).
\label{Gam4}
\end{equation}
Eq. (\ref{Gam4}) shows that at short times $\exp(-\Gamma(t))$ has a Gaussian shape with a width that depends on the structure and material parameters, as well as temperature. The inset of Fig. \ref{examplefree} compares the approximation of Eq. (\ref{Gam4}) with the the exact dephasing curve, Eq. (\ref{Gam1}).

\begin{figure}[t]
\includegraphics*[width=\linewidth]{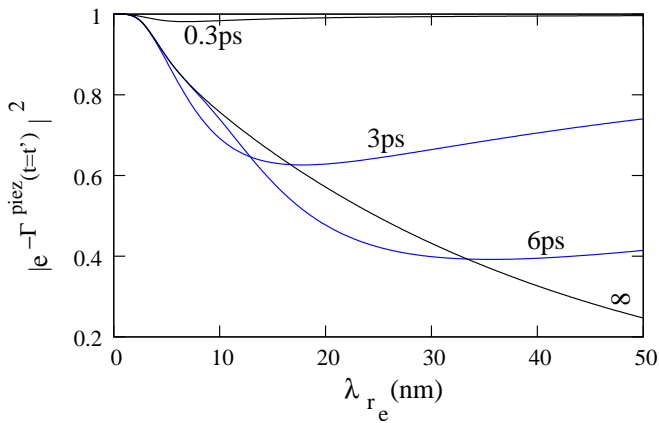}
\caption{Dependence of fixed-time dephasing due to piezoelectric coupling, 
$|\exp(-\Gamma^{piez}(t=t'))|^2$, with respect to $\lambda_{r_e}$ for an exciton at 77K. a) $t'=0.3$ ps, b) $t'=3$ ps, c) $t'=6$ ps.  In d), the 
long-time decay $|\exp(-\Gamma^{piez}(\infty))|^2$ is also depicted. }
\label{shorttpiez}
\end{figure}

For the piezoelectric interaction, the rate at which $\exp(-\Gamma^{piez}(t))$ falls from unity is temperature-dependent but for low temperatures the fastest dephasing (that is, the minimum width of the Gaussian in Eq. (\ref{Gam4})) occurs for $\lambda_{r_e}\approx\bar\lambda^{piez}_{r_e}$: at 4K, for instance, the fastest dephasing happens at $\lambda_{r_e}\approx6.18$~nm. 
For more oblate/prolate QDs, the initial dephasing occurs slower. Fig. \ref{shorttpiez} shows $|\exp(-\Gamma^{piez}(t=t'))|^2$ evaluated at different evolution times 
as a function of $\lambda_{r_e}$ and at $T$=77K. The short-time dephasing ($t'\lesssim 0.3$~ps) becomes negligible for small $\lambda_{r_e}$ as $I^{piez}\to0$, and also becomes negligible as $\lambda_{r_e}\to\infty$. Depending on $\lambda_{r_e}$, dephasing may be not monotonic with time. For example, in Fig. \ref{shorttpiez} $\exp(-\Gamma^{piez}(6$~ps$))<\exp(-\Gamma^{piez}(\infty))$ for $10$~nm $\lesssim \lambda_{r_e}\lesssim 33$~nm. This is due to the existence of a local minimum in the evolution of $\exp(-\Gamma^{piez}(t))$ before saturation occurs\cite{1stkuhn} (see Fig. \ref{examplefree}).

\begin{figure}[h]
\includegraphics*[width=\linewidth]{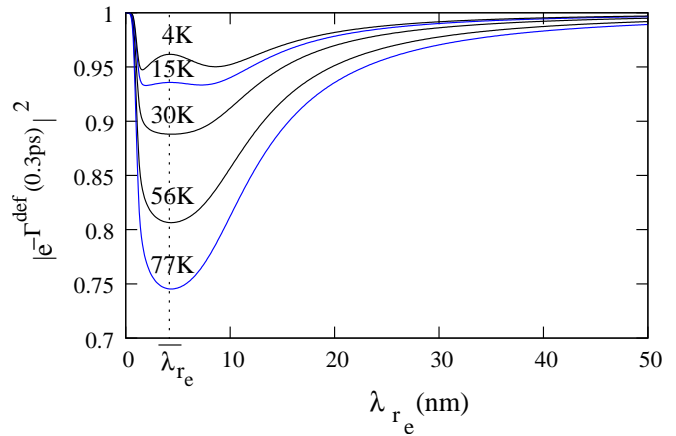}
\caption{Dependence of short-time dephasing due to deformation coupling, $|\exp(-\Gamma^{def}(t=0.3\mbox{ps}))|^2$, upon $\lambda_{r_e}$ for an exciton at different temperatures: a) 4K, b) 15K, c) 30K, d) 56K, and e) 77K.}
\label{shorttdef}
\end{figure}

For deformation potential coupling, the width of the Gaussian in Eq. (\ref{Gam4}) is locally maximum for QDs with $\lambda_{r_e}$ close to $\bar\lambda_{r_e}$. At 4K, this occurs at $\lambda_{r_e}\approx4.3$~nm. The short-time dephasing becomes negligible for very large and very small $\lambda_{r_e}$ (see Fig. \ref{shorttdef}, where $t=0.3$ ps). For low temperatures, a local minimum of the dephasing occurs near to $\bar\lambda_{r_e}$. As the temperature is increased, the local minimum is flattened and becomes a 
global maximum (see Fig. \ref{shorttdef}). 

\section{Periodically controlled exciton dephasing dynamics}

Having clarified the essential features of the phonon-induced dephasing dynamics in the absence of control, we now show how it can be substantially reduced through DD.  Our main goal here is to assess the benefits resulting from the application of 
DD in its simplest periodic form, involving repeated bit-flips as in Ref.~\onlinecite{lorenza}.

\subsection{Control setting}

Let DD be implemented by subjecting the exciton to a train of uniformly spaced ideal $\pi$-pulses, applied at instants $t_\ell = \ell \Delta t$, 
$\ell = 1, 2, \ldots$, with $\Delta t >0$ being the separation between consecutive pulses.  $T_c =2 \Delta t$ defines a complete DD cycle, which brings the control propagator back to unity.  The controlled dephasing dynamics may then be described by a modified 
decoherence function \cite{lorenza},
\begin{equation}
\Gamma (t)=\int_{0}^{\infty}\hspace*{-1mm}d\omega\frac{I(\omega)}{\omega^2}
\coth\Big(\frac{\hbar\omega}{k_B T}\Big)(1-\cos(\omega t))
\tan^2\Big(\frac{\omega\Delta t}{2}\Big), 
\label{Gam2}
\end{equation}
where time is understood to be stroboscopically sampled at integer multiples of the control cycle time, $t=N T_c= 2N\Delta t$. 

In practice, the required control rotations may be effected through suitable laser pulses.  On resonance, the exciton qubit will Rabi-oscillate between the logical states, and a pulse of appropriate length will perform a bit-flip. Eq. (\ref{Gam2}) assumes that such bit-flips are instantaneous, which is an adequate approximation provided the time necessary to perform each 
bit-flip operation is much shorter than any other time scale relevant to the problem.  In modeling the evolution of the excitonic qubit, however, a constraint must be placed on the timing of the control pulses. The central energy of a control pulse must be resonant with the energy of the ground state exciton, and its frequency spread must be small enough such that the probability of exciting higher energy levels is negligible. This places a lower limit on the pulse length, as too short a pulse would excite higher energy levels of the QD, and in extreme circumstances ionize the electron. To account for this in a model which assumes instantaneous bit-flips, the delay  between bit-flips must be significantly longer than the minimum length of each control pulse. 

The constraint on the pulse length is also a requirement in QD-based quantum computation schemes which are based on the so-called bi-excitonic shift~\cite{irene} to perform two-qubit gates. The bi-excitonic shift can be of the order of a few meV~\cite{irene}, thus the constraint translates into a pulse length of the order of a fraction of a picosecond.  The control pulses we are considering in this paper, however, can be shorter.  The relevant energy scale in this case is the difference between the ground state and higher energy levels, and in the strong confinement regime of interest here this can be of the order of tens of meV.  This translates to a minimum pulse length of the order of a few tens of femtoseconds, and a constraint on the interval between control pulses of the order of a few tenths of a picosecond. Finally, we assume that the lattice relaxation time is much greater than the interval between pulses, and that no systematic and/or random control errors affect the DD operations.

\subsection{Dephasing suppression through multipulse control}

\begin{figure}[t]
\includegraphics*[width=\linewidth]{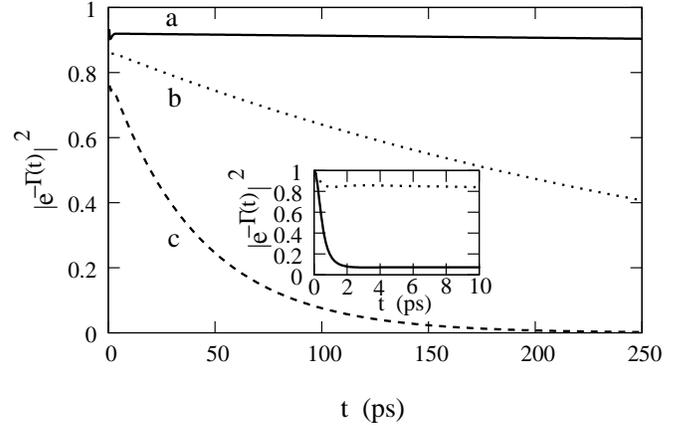}
\caption{Decoherence dynamics, $|\exp(-\Gamma(t))|^2$, for $\lambda_{r_e}=6.16$~nm at 77K under periodic DD with: a) $\Delta t=0.2$~ps, b) $\Delta t= 0.25$ ps, and c) $\Delta t= 0.3$~ps. Inset: free (solid line) and controlled (dotted line) evolution for the case of $\Delta t= 0.25$~ps. }
\label{examplecontrol}
\end{figure}

Figure \ref{examplecontrol} shows the decoherence factor, $|\exp(-\Gamma(t))|^2$, for $\lambda_{r_e}=6.16$~nm at 77K in the presence of a DD sequence 
with different pulse delays, whereas
the inset compares the free evolution with the controlled evolution on a smaller time scale for the case of $\Delta t= 0.25$~ps. Decreasing the pulse separation $\Delta t$ decreases the dephasing for two reasons: firstly, the qubit dephases slower in the presence of more closely separated control pulses, and secondly the control sequence begins at an earlier instant, 
thus less coherence is lost before the first control pulse occurs. 

As expected, the dephasing follows the free evolution until the first bit-flip occurs, after which a transient regime sets in during the first few control cycles. Eventually, the controlled coherence decays away to zero, but over a substantially longer time scale than the one associated with the free evolution.  Our numerical results suggest that no saturation at a non-zero value takes place for the controlled evolution in the relevant parameter regime, however the coherence decay time increases with decreasing $\Delta t$. This is apparent when comparing the controlled evolution over different time scales, e.g. compare the dotted lines in the main panel and the inset of Fig. \ref{examplecontrol}, respectively, which show the evolution for $\Delta t=0.25$~ps. As long as DD is turned on early enough, there exists a time interval in which the controlled evolution dephases less than the free evolution. This time interval can be orders of magnitude greater than the free evolution dephasing time.

Beside depending on the control time scale $\Delta t$, the DD efficiency is influenced by the shape of the dot, which enters Eq. (\ref{Gam2}) through the spectral density $I(\omega)$ analyzed in Sec. IIC.  Alternatively, one may regard the controlled decoherence behavior as being determined by a DD-renormalized spectral density $I(\omega)\tan^2(\omega\Delta t/2)$.  The last term presents a resonance at $\omega_{res}=\pi/\Delta t$, which is the characteristic frequency introduced by the periodic pulsing.  The smaller 
$\Delta t$, the higher this characteristic frequency is. Although the 
time-dependent contribution prevents the integrand from diverging to infinity, the location of $\omega_{res}$ relative to the bare spectral density plays a crucial role in determining the dephasing properties: at sufficiently long time $t$, the largest contribution to $\Gamma(t)$ for our system originates from frequencies close to $\omega_{res}$.  Therefore, DD is expected to be efficient provided that $\omega_{res} \gtrsim \omega_c$, i.e. physically, the control cycle time $T_c$ becomes significantly smaller than the correlation time scale $\tau_c=2\pi/ \omega_c$~\cite{remark}. 

\begin{figure}[t]
\includegraphics*[width=\linewidth]{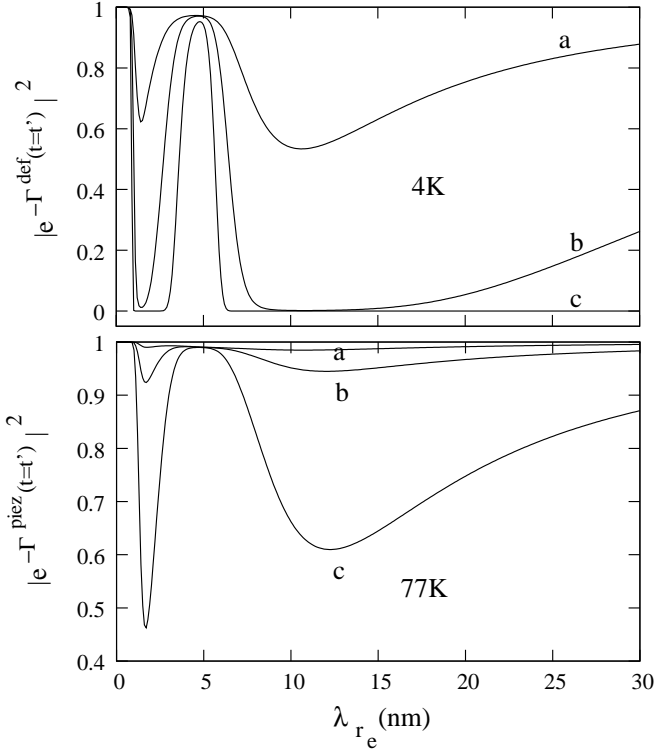}
\caption{Deformation decoherence factor $|\exp(-\Gamma^{def}(t'))|^2$ for an exciton at $4K$ (top panel) and piezoelectric decoherence factor $|\exp(-\Gamma^{piez}(t'))|^2$ for an exciton at $77$K (bottom panel), as a function of $\lambda_{r_e}$, in the presence of a sequence of DD pulses separated by $\Delta t= 0.3$~ps for different evolution times: a) $t'=3$~ps, b) $t'=30$~ps, and c) $t'=300$~ps.}
\label{pulsestdef}
\end{figure}

For control separations $\Delta t =0.3$~ps, $\omega_{res}=1.04$x$10^{13}$ rads$^{-1}$ ($\hbar\omega_{res}=6.85$~meV), which corresponds to frequencies 
$\omega \gtrsim\omega_c$ such that $I_{exc}(\omega)$ may become negligible depending on the shape of the dot (see Figs. \ref{deformexcitonI} and \ref{piezexcitonIlarger}).  In particular, the optimally shaped QD for DD 
is the one with the smallest $I_{exc}(\omega_{res})$. For $V_{\psi_e}=455.28$~nm$^3$ and $\Delta t=0.3$~ps, this yields $\lambda_{r_e}=4.8$~nm.
The top panel of Fig. \ref{pulsestdef} depicts the decoherence factor due to deformation coupling, $|\exp(-\Gamma^{def}(t'))|^2$, for an exciton at $4$K subject to a DD sequence with $\Delta t=0.3$~ps evaluated at 
various evolution times $t'$. 
Full dephasing may occur for very oblate/prolate QDs, however there is also a local maximum in the efficiency of the control sequence, which for large $t'$ occurs for dots with $\lambda_{r_e}=4.8$~nm. This is due to this particular shape of dot having the smallest $I^{def}_{exc}(\omega_{res})$. The bottom panel of Fig. \ref{pulsestdef} shows the piezoelectric contribution, $|\exp(-\Gamma^{piez}(t'))|^2$, for an exciton at 77K in the same control settings.
Once again, the local maximum in the DD efficiency for large $t'$ is determined by the $\lambda_{r_e}$ which minimizes $I^{piez}_{exc}(\omega_{res})$. At lower temperatures, the overall magnitude of the dephasing is reduced.

\begin{figure}[t]
\includegraphics*[width=\linewidth]{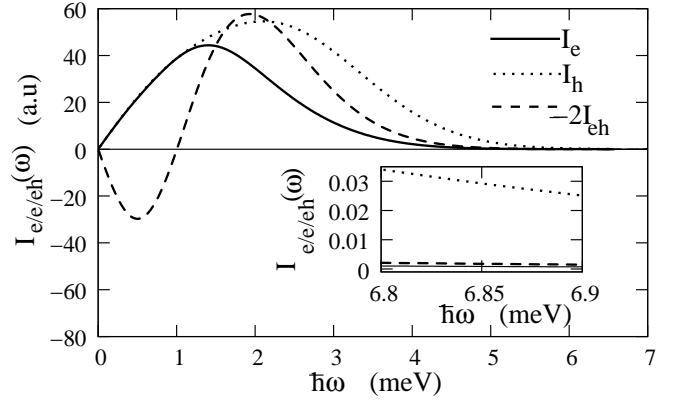}
\caption{Spectral density contributions $I_e(\omega)$, $I_h(\omega)$, and $-2I_{eh}(\omega)$ for a QD with $\lambda_{r_e}=4.8$~nm. Inset: $I_e(\omega)$, $I_h(\omega)$, and $-2I_{eh}(\omega)$ at $\omega=\omega_{res}=10.4$x$10^{12}$~rads$^{-1}$.}
\label{components}
\end{figure}

The separate components of the spectral density $I_e$, $I_h$, and $-2I_{eh}$ for a QD with $\lambda_{r_e}=4.8$~nm (corresponding to the minimum dephasing in the top panel of Fig. \ref{pulsestdef}) are plotted in Fig. \ref{components}. The components of the spectral density include contributions from both piezoelectric and deformation potential coupling. For this reason, the interference term $I_{eh}$ includes both negative and positive terms. It can be seen that $I_{exc}\approx I_{h}$ for $\omega \gtrsim 8$x$10^{12}$~rads$^{-1}$. Therefore, close to the local minimum of the dephasing we can identify the dot shape needed to minimize the dephasing by choosing $\lambda_{r_e}$ such that $I_h$ is minimized at high $\omega$.  This occurs when the hole wavefunction is spherical. For $V_{\psi_e}=455.28$~nm$^3$ and GaAs 
(i.e. $\lambda_{r_h}=0.495\lambda_{r_e}$), this corresponds to $\lambda_{r_e}=4.856$~nm, which is consistent with the minimum dephasing value of $\lambda_{r_e}=4.8$~nm. The small difference is due to the contributions to $I_{exc}(\omega_{res})$ from $I_{eh}$ and $I_e$, see inset of Fig. \ref{components}.

Note that, as expected, as the limit of very fast control is approached ($T_c \ll \tau_c$), the shape of the dot has a smaller influence on the exciton coherence, which generally remains close to its value at the instant the first pulse occurred for a long period of time. 

\subsection{DD performance for exciton qubits}

\begin{figure}[h]
\includegraphics*[width=\linewidth]{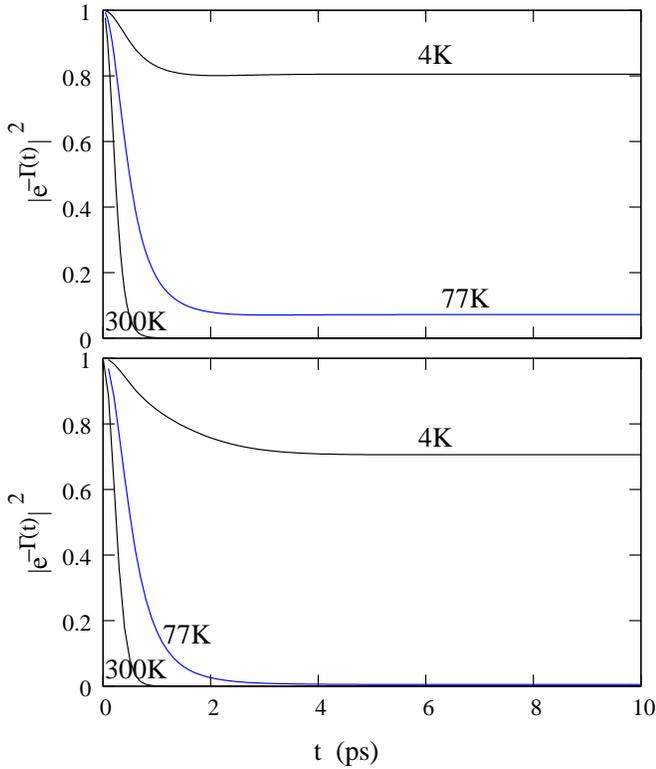}
\caption{Free dephasing dynamics of QD A (upper panel), and QD B (lower panel) for different temperatures.}
\label{rhonat}
\end{figure}

Let us now specifically focus on QDs intended for quantum computation applications~\cite{irene}.  We will consider two sets of parameters, QD A and QD B. QD A is a GaAs QD with an AlAs barrier. The confining potentials are modeled as parabolic in all three dimensions, with electron (hole) confinement energies in the $z$-direction of $\hbar\omega_e=505$~meV ($\hbar\omega_h=100$~meV), and $\hbar\omega_e=30$~meV 
($\hbar\omega_h=24$~meV) in the in-plane directions, respectively. This corresponds to $\lambda_{r_e}=6.16$~nm, $\lambda_{r_h}=3.05$~nm, and 
$\lambda_{z_e}=\lambda_{z_h}=1.5$~nm. QD B is the same dot, except an 
external electric field of $75$~kV/cm is applied in the $x$-direction.  
QD B has been shown to be suitable for quantum information processing schemes which exploit direct Coulomb coupling between excitons~\cite{irene}, whereas QD A may be useful in schemes not based on the biexcitonic shift. The time evolution of the off-diagonal density matrix elements for the free dephasing dynamics (Eq. (\ref{dens})) for QDs A and B is depicted in Fig. 
\ref{rhonat} for different temperatures~\cite{kuhnirene}. 

\begin{figure}
\includegraphics*[width=\linewidth]{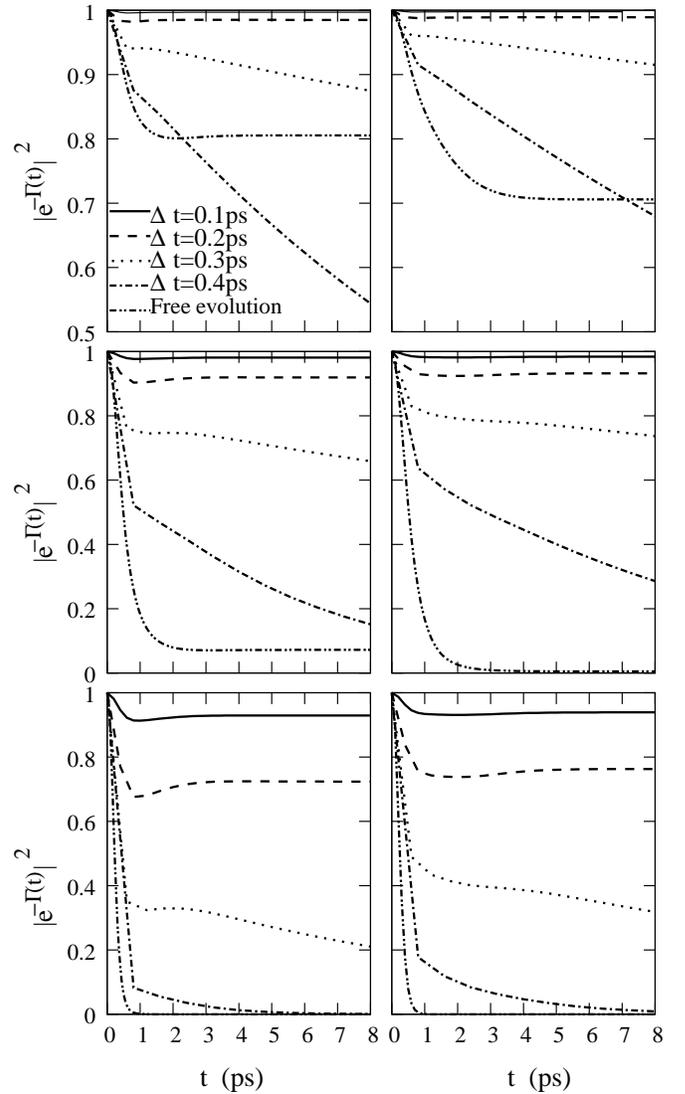}
\caption{Controlled dephasing dynamics of QD A (left), and QD B(right) at 
4K (top), 77K (middle), and 300K (bottom), for different pulse separations.}\label{computpulses}
\end{figure}

Figure \ref{computpulses} shows the qubit coherence evolution for both QDs in the presence of periodic DD, according to Eq. (\ref{Gam1}). The left column corresponds to QD A, and the right to QD B. Temperatures are 4K, 77K, and 300K from top to bottom. It can be seen that for pulses separated by $\Delta t \lesssim 0.2$~ps, DD approximately freezes the dephasing over the time scales relevant to the problem. For a control sequence with longer pulse delays, the qubit coherence decays more rapidly, however dephasing can still be strongly suppressed for relatively long times.

In terms of practical quantum computation, the bit-flips do not change the time taken for the onset of decoherence, that is, the initial time taken for $e^{-\Gamma(t)}$ to sharply decrease from unity. The qubit follows its free evolution until the first bit-flip occurs, which due to the physical constraint on $\Delta t$, happens after some coherence is lost. Still, the pulse sequence succeeds at significantly enhancing the exciton coherence at large $t$. For certain operating temperatures and/or QD devices, the resulting coherence level may remain very close to unity, e.g. for QD B at $4$K and $\Delta t =0.1$~ps, $\rho_{01}(t=10$~ps)$=\sqrt{0.997}\rho_{01}(t=0)$, to be contrasted with the free evolution value of $\sqrt{0.706}\rho_{01}(t=0)$ (see right upper panel). Even at room temperature, where for the free evolution all coherence is lost in the first picosecond, our calculations show that for $\Delta t=0.1$~ps, $\rho_{01}(t) > \sqrt{0.929}\rho_{01}(t=0)$ for both QD A and QD B for a long period of time. 

\begin{figure}[t]
\includegraphics*[width=\linewidth]{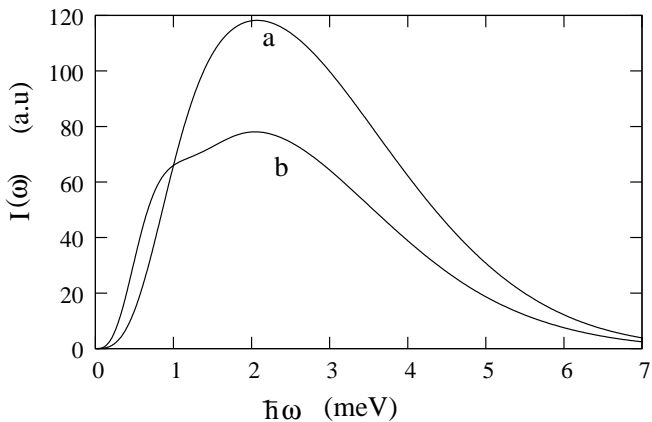}
\caption{Spectral density $I(\omega)$ for a) exciton in QD A, and 
b) exciton in QD B.}\label{spectcomparison}
\end{figure}

It is important to note that in the presence of DD, the qubit subject to the applied electric field (QD B) dephases {\em less} than without applied field (compare left and right column of Fig. \ref{computpulses}). This is in contrast to the free evolution behavior, whereby the presence of an electric field increases the coupling to the piezoelectric field and enhances the dephasing. However, a static electric field increases the coupling between excitonic qubits and significantly improves the feasibility of quantum computation. Our results demonstrate that an electric field actually {\em increases} the efficiency of periodic DD.

This improvement 
may be understood by looking at the exciton spectral density. Fig. \ref{spectcomparison} compares the spectral density for the exciton with and without electric field. It can be seen that the spectral density in the presence of an electric field is shifted to {\em lower} frequencies compared to the no field case. This leads to larger long-term dephasing (see Eq. (\ref{Gam3})), in a way  similar to the increase of dephasing due to the piezoelectric field as $\lambda_{r_e}$ increases. Introducing the electric field effectively causes the piezoelectric coupling to contribute 
more significantly to the dephasing than the deformation potential coupling~\cite{kuhnirene}, due to the non-vanishing electron-hole dipole moment. Because the relevant spectral density is proportional to the Fourier transform of the wave-function, the relative shift of the electron and hole wave-functions results in a narrower spectral density.  In particular, it  decreases the spectral density at high frequencies (see Fig. \ref{spectcomparison}). As a result, the efficiency of sufficiently fast DD sequences ($\omega_c \Delta t \lesssim \pi$) is increased due to a lower contribution at $\omega_{res}$, as discussed in Sec. III.B. More generally, any phenomenon which decreases the spectral density at $\omega_{res}$ 
will automatically improve DD performance. 

Also note that for very short pulse separations ($\omega_c \Delta t \ll \pi$), coherence is maintained near its value at the instant of the first control pulse.  Since the applied electric field leads to a slower initial dephasing than the no field case, less coherence has been lost at the instant of the first control pulse, thus $\rho_{01}$ remains at a higher value. Again, this is similar to the piezoelectric dephasing behavior observed for increasing $\lambda_{r_e}$.

Effectively, the presence of an electric field or a QD shape such that 
$\lambda_{r_e}$ is close to $\bar\lambda_{r_e}$, $\bar\lambda^{piez}_{r_e}$ leads to greater dephasing, both in terms of the speed of the initial dephasing, as well as the magnitude at which coherence saturates, as seen in Sec. IIC. However, our results show that it is precisely for these QDs that the DD scheme is most efficient, for the shortest control sequences allowed by the physical system. 
Specifically, the QDs described in this section have $\lambda_{r_e}=6.16$~nm, which is close to that for which DD is most efficient (see Fig. \ref{pulsestdef}). Slightly changing the dot dimensions so that $\lambda_{r_e}$ becomes closer to the ideal value would further improve the results reported in Fig. \ref{computpulses}.

\subsection{DD for short-term quantum storage}

Our calculations indicate that DD may effectively decouple the exciton qubit from the environment for a long time, allowing robust storage of quantum information. In standard solid-state quantum computation schemes, holes in double dots \cite{storage} and spin qubits \cite{paulibl} are employed for long-term quantum information storage, due to longer $T_1$ lifetimes, whilst exciton qubits are usually proposed for computation only, often as ancillary states, and used for gating purposes. However, quantum algorithms are generally built up of a series of gates, and {\em arbitrary} qubit states may have to be stored {\it temporarily} while other gate operations are carried out.  We propose that DD be invoked
to achieve robust {\em short-term storage}. 

\begin{figure}[t]
\includegraphics*[width=\linewidth]{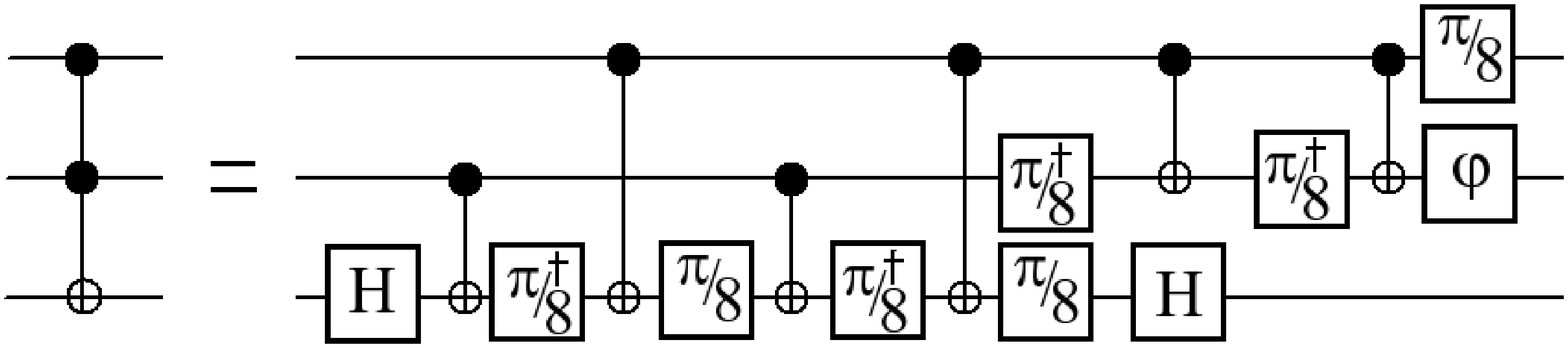}
\caption{A simple Toffoli gate designed from six controlled-not gates, six 
$\pi/8$-gates, two Hadamard, and a phase gate.}
\label{toffoli}
\end{figure}

As a concrete example, consider the Toffoli gate\cite{neilsenchang} given in Fig. \ref{toffoli}. 
While this is a very simple three-qubit operation, it still requires the first control qubit to be stored for the time taken to perform 8 gate operations. If we consider an average gating time of $\sim 1$ ps per gate, Fig. \ref{rhonat} shows that unless techniques like the one we propose are used, substantial decoherence would occur on these timescales. 

\section{Conclusions}

We have shown that the dephasing of an exciton qubit due to its coupling with the lattice phonons can be significantly decreased by coherent dynamical control. We found that whilst quantum dots with a size and shape most suitable for quantum computation are the most prone to dephasing, periodic dynamical decoupling is most efficient for exactly this type of dot. We have demonstrated that a good estimate for the shape of the dot which optimizes decoupling performance is the one where the hole wavefunction is spherical.  In addition, our analysis indicates that, although the presence of an electric field (which is required by most exciton-based quantum computing schemes) increases dephasing, it also makes for improved decoupling performance. In the presence of periodic decoupling, the decay of qubit coherence is substantially reduced over evolution times which are very long compared to typical gating times.  Under favorable circumstances, it is reduced so significantly that in practical terms the qubit no longer dephases. We propose that dynamical decoupling in the simplest periodic implementation examined here may be useful to improve temporary data storage during the memory stage of quantum computation. 

Besides assessing the impact of pulse imperfections which are unavoidably present in real control systems, several extensions 
of the present analysis may be worth considering.  In particular, dynamical decoupling schemes which involve non-uniform time delays and are based on either recursive concatenated design \cite{kaveh} or direct cancellation of high-order error terms \cite{uhrig} have been recently found to attain remarkable high decoupling fidelity under appropriate assumptions \cite{kaveh,wen,witzel2}.  Furthermore, so-called Eulerian decoupling schemes \cite{euler} allow decoherence suppression to be achieved without requiring unrealistic control strengths as in the bang-bang limit -- which may be critical to minimize unwanted excitations.  A future investigation will be needed to identify the possible added benefits of more elaborated decoupling schemes under the specific physical constraints and design trade-offs associated with exciton based quantum-dot devices. 

\acknowledgments

I.D. acknowledges financial support from the Department of Physics of the University of York and the kind hospitality of the Department of Physics and
Astronomy of Dartmouth College during the early stages of this work.  L.V. also gratefully acknowledges partial support from the NSF through Grant No. PHY-0555417, and from the Department of Energy, Basic Energy Sciences, under Contract No. DE-AC02-07CH11358.

\end{document}